\documentstyle[twoside]{article}

\catcode`\@=11
\long\def\@makefntext#1{
\protect\noindent \hbox to 3.2pt {\hskip-.9pt  

$^{{\eightrm\@thefnmark}}$\hfil}#1\hfill}		%CAN BE USED 

\def\thefootnote{\fnsymbol{footnote}}
\def\@makefnmark{\hbox to 0pt{$^{\@thefnmark}$\hss}}	%ORIGINAL 

\def\ps@myheadings{\let\@mkboth\@gobbletwo
\def\@oddhead{\hbox{}
\rightmark\hfil\eightrm\thepage}   

\def\@oddfoot{}\def\@evenhead{\eightrm\thepage\hfil
\leftmark\hbox{}}\def\@evenfoot{}
\def\sectionmark##1{}\def\subsectionmark##1{}}

\oddsidemargin=\evensidemargin
\renewcommand{\thefootnote}{\fnsymbol{footnote}}

\newcounter{sectionc}\newcounter{subsectionc}\newcounter{subsubsectionc}
\renewcommand{\section}[1] {\vspace{12pt}\addtocounter{sectionc}{1} 

\setcounter{subsectionc}{0}\setcounter{subsubsectionc}{0}\noindent 

	{\tenbf\thesectionc. #1}\par\vspace{5pt}}
\renewcommand{\subsection}[1] {\vspace{12pt}\addtocounter{subsectionc}{1} 

	\setcounter{subsubsectionc}{0}\noindent 

	{\bf\thesectionc.\thesubsectionc. {\kern1pt \bfit #1}}\par\vspace{5pt}}
\renewcommand{\subsubsection}[1] {\vspace{12pt}\addtocounter{subsubsectionc}{1}
	\noindent{\tenrm\thesectionc.\thesubsectionc.\thesubsubsectionc.
	{\kern1pt \tenit #1}}\par\vspace{5pt}}
\newcommand{\nonumsection}[1] {\vspace{12pt}\noindent{\tenbf #1}
	\par\vspace{5pt}}

\newcounter{appendixc}
\newcounter{subappendixc}[appendixc]
\newcounter{subsubappendixc}[subappendixc]
\renewcommand{\thesubappendixc}{\Alph{appendixc}.\arabic{subappendixc}}
\renewcommand{\thesubsubappendixc}
	{\Alph{appendixc}.\arabic{subappendixc}.\arabic{subsubappendixc}}

\renewcommand{\appendix}[1] {\vspace{12pt}
        \refstepcounter{appendixc}
        \setcounter{figure}{0}
        \setcounter{table}{0}
        \setcounter{lemma}{0}
        \setcounter{theorem}{0}
        \setcounter{corollary}{0}
        \setcounter{definition}{0}
        \setcounter{equation}{0}
        \renewcommand{\thefigure}{\Alph{appendixc}.\arabic{figure}}
        \renewcommand{\thetable}{\Alph{appendixc}.\arabic{table}}
        \renewcommand{\theappendixc}{\Alph{appendixc}}
        \renewcommand{\thelemma}{\Alph{appendixc}.\arabic{lemma}}
        \renewcommand{\thetheorem}{\Alph{appendixc}.\arabic{theorem}}
        \renewcommand{\thedefinition}{\Alph{appendixc}.\arabic{definition}}
        \renewcommand{\thecorollary}{\Alph{appendixc}.\arabic{corollary}}
        \renewcommand{\theequation}{\Alph{appendixc}.\arabic{equation}}
        \noindent{\tenbf Appendix \theappendixc #1}\par\vspace{5pt}}
\newcommand{\subappendix}[1] {\vspace{12pt}
        \refstepcounter{subappendixc}
        \noindent{\bf Appendix \thesubappendixc. {\kern1pt \bfit #1}}
	\par\vspace{5pt}}
\newcommand{\subsubappendix}[1] {\vspace{12pt}
        \refstepcounter{subsubappendixc}
        \noindent{\rm Appendix \thesubsubappendixc. {\kern1pt \tenit #1}}
	\par\vspace{5pt}}

\newcommand{\textlineskip}{\baselineskip=13pt}
\newcommand{\smalllineskip}{\baselineskip=10pt}

\def\eightcirc{
\begin{picture}(0,0)
\put(4.4,1.8){\circle{6.5}}
\end{picture}}
\def\eightcopyright{\eightcirc\kern2.7pt\hbox{\eightrm c}}

\newcommand{\copyrightheading}[1]
	{\vspace*{-2.5cm}\smalllineskip{\flushleft
	{\footnotesize International Journal of Modern Physics A, #1}\\
	{\footnotesize $\eightcopyright$\, World Scientific Publishing
	 Company}\\
	 }}

\def\abstracts#1#2#3{{
	\centering{\begin{minipage}{4.5in}\baselineskip=10pt\footnotesize
	\parindent=0pt #1\par 

	\parindent=15pt #2\par
	\parindent=15pt #3
	\end{minipage}}\par}}

\renewenvironment{thebibliography}[1]
	{\frenchspacing
	 \ninerm\baselineskip=11pt
	 \begin{list}{\arabic{enumi}.}
	{\usecounter{enumi}\setlength{\parsep}{0pt}
	 \setlength{\leftmargin 12.7pt}{\rightmargin 0pt} %FOR 1--9 ITEMS
	 \setlength{\itemsep}{0pt} \settowidth
	{\labelwidth}{#1.}\sloppy}}{\end{list}}

\newcounter{itemlistc}
\newcounter{romanlistc}
\newcounter{alphlistc}
\newcounter{arabiclistc}

\newcommand{\fcaption}[1]{
        \refstepcounter{figure}
        \setbox\@tempboxa = \hbox{\footnotesize Fig.~\thefigure. #1}
        \ifdim \wd\@tempboxa > 5in
           {\begin{center}
        \parbox{5in}{\footnotesize\smalllineskip Fig.~\thefigure. #1}
            \end{center}}
        \else
             {\begin{center}
             {\footnotesize Fig.~\thefigure. #1}
              \end{center}}
        \fi}

\newcommand{\tcaption}[1]{
        \refstepcounter{table}
        \setbox\@tempboxa = \hbox{\footnotesize Table~\thetable. #1}
        \ifdim \wd\@tempboxa > 5in
           {\begin{center}
        \parbox{5in}{\footnotesize\smalllineskip Table~\thetable. #1}
            \end{center}}
        \else
             {\begin{center}
             {\footnotesize Table~\thetable. #1}
              \end{center}}
        \fi}

\def\@citex[#1]#2{\if@filesw\immediate\write\@auxout
	{\string\citation{#2}}\fi
\def\@citea{}\@cite{\@for\@citeb:=#2\do
	{\@citea\def\@citea{,}\@ifundefined
	{b@\@citeb}{{\bf ?}\@warning
	{Citation `\@citeb' on page \thepage \space undefined}}
	{\csname b@\@citeb\endcsname}}}{#1}}

\newif\if@cghi
\def\cite{\@cghitrue\@ifnextchar [{\@tempswatrue
	\@citex}{\@tempswafalse\@citex[]}}
\def\citelow{\@cghifalse\@ifnextchar [{\@tempswatrue
	\@citex}{\@tempswafalse\@citex[]}}
\def\@cite#1#2{{$\null^{#1}$\if@tempswa\typeout
	{IJCGA warning: optional citation argument 

	ignored: `#2'} \fi}}

\def\pmb#1{\setbox0=\hbox{#1}
	\kern-.025em\copy0\kern-\wd0
	\kern.05em\copy0\kern-\wd0
	\kern-.025em\raise.0433em\box0}

\def\fnt#1#2{\footnotetext{\kern-.3em
	{$^{\mbox{\scriptsize #1}}$}{#2}}}

\def\fpage#1{\begingroup
\voffset=.3in
\thispagestyle{empty}\begin{table}[b]\centerline{\footnotesize #1}
	\end{table}\endgroup}

\def\runninghead#1#2{\pagestyle{myheadings}
\markboth{{\protect\footnotesize\it{\quad #1}}\hfill}
{\hfill{\protect\footnotesize\it{#2\quad}}}}
\font\tenrm=cmr10
\font\tenit=cmti10 

\font\tenbf=cmbx10
\font\bfit=cmbxti10 at 10pt
\font\ninerm=cmr9

\font\eightrm=cmr8

\def\qed{\hbox{${\vcenter{\vbox{			%HOLLOW SQUARE
   \hrule height 0.4pt\hbox{\vrule width 0.4pt height 6pt
   \kern5pt\vrule width 0.4pt}\hrule height 0.4pt}}}$}}

\renewcommand{\thefootnote}{\fnsymbol{footnote}}	%USE SYMBOLIC FOOTNOTE
\def\beqra{\begin{eqnarray}} \def\eeqra{\end{eqnarray}}
\def\beqast{\begin{eqnarray*}} \def\eeqast{\end{eqnarray*}}
\def\beq{\begin{equation}}      \def\eeq{\end{equation}}
\def\be{\begin{enumerate}}   \def\ee{\end{enumerate}}

\def\gam{\gamma}

\def\la{\lambda}

\def\si{\sigma}

\def\del{\delta}

\def\om{\omega}

\def\pa{\partial}

\def\cd{{\cal D}}

\def\cz{{\cal{Z}}}

\def\raisenot{\raise .5mm\hbox{/}}
\def\nota{\ \hbox{{$a$}\kern-.49em\hbox{/}}}
\def\notA{\hbox{{$A$}\kern-.54em\hbox{\raisenot}}}
\def\notb{\ \hbox{{$b$}\kern-.47em\hbox{/}}}
\def\notB{\ \hbox{{$B$}\kern-.60em\hbox{\raisenot}}}
\def\notc{\ \hbox{{$c$}\kern-.45em\hbox{/}}}
\def\notd{\ \hbox{{$d$}\kern-.53em\hbox{/}}}
\def\notbd{\ \hbox{{$D$}\kern-.61em\hbox{\raisenot}}} %big D
\def\note{\ \hbox{{$e$}\kern-.47em\hbox{/}}}
\def\notk{\ \hbox{{$k$}\kern-.51em\hbox{/}}}
\def\notp{\ \hbox{{$p$}\kern-.43em\hbox{/}}}
\def\notq{\ \hbox{{$q$}\kern-.47em\hbox{/}}}
\def\notW{\ \hbox{{$W$}\kern-.75em\hbox{\raisenot}}}
\def\notz{\ \hbox{{$Z$}\kern-.61em\hbox{\raisenot}}}
\def\notpa{\hbox{{$\partial$}\kern-.54em\hbox{\raisenot}}}

\def\fo{\hbox{{1}\kern-.25em\hbox{l}}}  %raised one

\def\tr{{\rm Tr}}

\def\lag{\langle}
\def\rag{\rangle}

\def\asymptotic{{_{\stackrel{\displaystyle\longrightarrow}
{x\rightarrow\pm\infty}}\,\, }} %x goes to plus minus infinity, display sty.
\def\asymptext{\raisebox{.6ex}{${_{\stackrel{\displaystyle\longrightarrow}
{x\rightarrow\pm\infty}}\,\, }$}} %x goes to plus minus infinity, within text.

\def\7#1#2{\mathop{\null#2}\limits^{#1}}        % puts #1 atop #2
\def\5#1#2{\mathop{\null#2}\limits_{#1}}        % puts #1 beneath #2

\def\place#1#2#3{\vbox to0pt{\kern-\parskip\kern-7pt
                             \kern-#2truein\hbox{\kern#1truein #3}
                             \vss}\nointerlineskip}

\def\illustration #1 by #2 (#3){\vbox to #2{\hrule width #1 height 0pt 
depth
0pt
                                       \vfill\special{illustration #3}}}

\def\scaledillustration #1 by #2 (#3 scaled #4){{\dimen0=#1 \dimen1=#2
           \divide\dimen0 by 1000 \multiply\dimen0 by #4
            \divide\dimen1 by 1000 \multiply\dimen1 by #4
            \illustration \dimen0 by \dimen1 (#3 scaled #4)}}

\def\sigx{\sigma(x)}
\def\pix{\pi(x)}

\def\pax{\pa_x}

\renewcommand{\theequation}{\thesection.\arabic{equation}}

\begin{document}

\runninghead{Dynamical Generation of Solitons  $\dots$}  
{J. Feinberg \& A. Zee} 

\normalsize\textlineskip
\thispagestyle{empty}
\setcounter{page}{1}

\copyrightheading{}			%{Vol. 0, No. 0 (1993) 000--000}

\vspace*{0.88truein}

\fpage{1}
\centerline{\bf DYNAMICAL GENERATION OF SOLITONS}
\vspace*{0.035truein}
\centerline{\bf IN A $1+1$ DIMENSIONAL CHIRAL FIELD THEORY:}
\vspace*{0.035truein}
\centerline{\bf NON-PERTURBATIVE DIRAC OPERATOR RESOLVENT ANALYSIS
\footnote{Talk delivered by J.Feinberg at the Workshop on Low Dimensional Field Theory at Telluride, CO (August 1996). To appear in the conference proceedings. }}
\vspace*{0.37truein}
\centerline{\footnotesize JOSHUA FEINBERG \& A. ZEE}
\vspace*{0.015truein}
\centerline{\footnotesize\it Institute for Theoretical Physics, }
\baselineskip=10pt
\centerline{\footnotesize\it University of California, Santa Barbara, CA 93106, USA}
\vspace*{10pt}
\vspace*{0.225truein}
%\publisher{August 4, 1996}

\vspace*{0.21truein}
\abstracts{We analyze the $1+1$ dimensional Nambu-Jona-Lasinio model 
non-perturbatively. We study non-trivial saddle points of the effective action in which the composite fields $\sigx=\lag\bar\psi\psi\rag$ and $\pix=\lag\bar\psi i\gam_5\psi\rag$ form static space dependent configurations. These configurations may be viewed as one dimensional chiral bags that trap the original fermions (``quarks") into stable extended entities (``hadrons"). We provide explicit expressions for the profiles of some of these objects and calculate their masses. Our analysis of these saddle points, and in particular, 
the proof that the $\sigx, \pix$ condensations must give rise to a reflectionless Dirac operator, appear to us simpler and more direct than the calculations previously done by Shei, using 
the inverse scattering method following Dashen, Hasslacher, and Neveu. }{}{}

%\textlineskip			%) USE THIS MEASUREMENT WHEN THERE IS
%\vspace*{12pt}			%) NO SECTION HEADING

\vspace*{1pt}\textlineskip	%) USE THIS MEASUREMENT WHEN THERE IS
\section{Introduction}
\setcounter{equation}{0}
\renewcommand{\theequation}{1.\arabic{equation}}

	%) A SECTION HEADING
\vspace*{-0.5pt}
\noindent
In this talk we describe a novel method\cite{feinzee}  for studying the  non-perturbative spectrum of the $1+1$ dimensional Nambu-Jona-Lasinio (NJL) model\cite{njl,gn}
\beq
S=\int d^2x\left\{\sum_{a=1}^N\, \bar\psi_a\,i\notpa\,\psi_a +
\frac{g^2}{2}\; \left[\left( \sum_{a=1}^N\;\bar\psi_a\,\psi_a\right)^2
-\left(\sum_{a=1}^N\;\bar\psi_a\gam_5\psi_a\right)^2\right]\right\}\,.
\label{lagrangian}
\eeq
The action (\ref{lagrangian}) describes $N$ self interacting massless Dirac fermions $\psi_a\,(a=1,\ldots,N)$, and we study it in the limit $N\rightarrow\infty$ holding $Ng^2$ finite (the large $N$ limit). This model is interesting because it shares with QCD some of its important low energy attributes\cite{chirallagrangian}, and at the same time its non-trivial dynamics is under control. Indeed, (\ref{lagrangian}) is invariant under $SU(N)_f\otimes U(1)\otimes U(1)_A$. It is asymptotically free and exhibits dynamical mass generation (for $N\geq 2$) due to its infra-red instabilities. These instabilities may also polarize the vacuum inhomogenously, giving rise to chiral solitons\cite{shei,feinzee} that are reminiscent of chiral bags\cite{sphericalbag,shellbag} which model hadron formation. These chiral solitons and the small fermionic fluctuations around them, are the main subject of this talk.  We rewrite (\ref{lagrangian}) as
\beq
S=\int 
d^2x\,\left\{\bar\psi\left[i\notpa-(\si+i\pi\gam_5)\right]\psi-{1\over 
2g^2}\,(\si^2+\pi^2)\right\}
\label{auxiliary}
\eeq
where $\si(x), \pi(x)$ are the scalar and pseudoscalar auxiliary fields, 
respectively,\footnote{From this point on flavor indices
are suppressed.} which are both of mass dimension $1$. These 
fields are singlets under $SU(N)_f\otimes U(1)$, but transform as a vector
under the axial transformation, namely $\si+i\gam_5\pi\rightarrow e^{-2i\gam_5\beta}(\si + i\gam_5\pi)\,$. Gaussian integration over the fermions in (\ref{auxiliary}) leads to the partition function $\cz=\int\,\cd\si\,\cd\pi\,\exp \{iS_{eff}[\si,\pi]\}$
where the bare effective action is
\beq
S_{eff}[\si,\pi] =-{1\over 2g^2}\int\, d^2x 
\,\left(\si^2+\pi^2\right)-iN\, 
\tr\ln\left[i\notpa-\left(\si+i\pi\gam_5\right)\right]
\label{effective}
\eeq
and the trace is taken over both functional and Dirac indices. 

The non-perturbative vacuum of (\ref{effective}) is determined, in the large  $N$ limit\cite{gn}, by the simplest large $N$ saddle points of (\ref{effective}) where $\si$ and $\pi$ develop spacetime independent expectation values. These saddle points are extrema of the effective potential $V_{eff}$, which as a result of chiral symmetry depends only on the combination $\rho^2=\si^2+\pi^2$. It is minimized at $\rho = m \Lambda\,e^{-{\pi\over Ng^2\left(\Lambda\right)}}$, where $m$ is the dynamical mass of fermions. This mass is determined by the (bare) gap equation\cite{gn}
\beq
-m + iNg^2\,{\rm tr}\int
{d^2k\over\left(2\pi\right)^2}{1\over\notk-m}
= 0\,,
\label{bgap}
\eeq
were $ \Lambda$ is an ultraviolet cutoff. The mass $m$ must be a 
renormalization group invariant. Thus, the model is asymptotically 
free. The vacuum manifold of (\ref{auxiliary}) is therefore a circle 
$\rho=m$ in the $\si,\pi$ plane, and the equivalent vacua are 
parametrized by the chiral angle $\theta={\rm arctan} {\pi\over\sigma}$. Therefore, small fluctuations of the Dirac fields around the vacuum manifold 
develop dynamical\footnote{Note that the axial 
$U(1)$ symmetry protects the fermions from 
developing a mass term to any order in perturbation theory.} 
chiral mass $m\,{\rm exp} (i\theta \gam_5)$. \footnote{Note in passing that the massless fluctuations of $\theta$ along the vacuum manifold decouple from the spectrum \cite{decouple} so that the axial $U(1)$ symmetry does not break dynamically in this two dimensional model \cite{coleman}.}

Static, space dependent solutions of the saddle point equations of (\ref{effective})
\beqra
{\del S_{\em eff}\over \del \si\left(x,t\right)}  &=&
-{\si\left(x,t\right)\over g^2} + iN ~{\rm tr} \left[~~~~~ \langle x,t | 
{1\over i\notpa
-\left(\si + i\pi\gam_5\right)} | x,t \rangle \right]= 0
\nonumber\\{}\nonumber\\
{\del S_{\em eff}\over \del \pi\left(x,t\right)}  &=&
-{\pi\left(x,t\right)\over g^2} - ~N~ {\rm tr} \left[~\gam_5~\langle x,t | {1\over i\notpa
-\left(\si + i\pi\gam_5\right)} | x,t \rangle~\right] = 0
\label{saddle}
\eeqra
are the chiral solitons (at rest) mentioned above.
The NJL model, with its continuous symmetry, does not have topologically stable soliton solutions. The  solitons arising in the NJL model can only be stabilized by binding fermions and releasing binding energy\cite{mackenzie}. This observation is clarified further by comparing the NJL model to the Gross-Neveu\cite{gn} (GN) model, namely, (\ref{lagrangian}) with the $\gam_5$
term deleted. The GN model possesses only a discrete symmetry, $\psi\rightarrow \gam_5\psi$, rather than the continuous symmetry of the NJL model. This discrete symmetry is dynamically broken by the doubly degenerate non-perturbative vacuum, and thus there is a topologicaly stable kink solution\cite{ccgz,dhn,josh1}, the so-called Callan-Coleman-Gross-Zee (CCGZ) kink, that interpolates between the two vacua. The kink binds any number $n\leq N$ of fermions in its single zero energy bound state, without affecting its mass. Its stability is guaranteed by topology already. In contrast, the stability of the extended objects arising in the NJL model is not due to topology, but to dynamics.

Shei \cite{shei} has found certain static solutions of (\ref{saddle})
by applying the inverse scattering method\cite{faddeev}, following a similar analysis by Dashen et al.\cite{dhn}. Recently, one of us developed an alternative method\cite{josh1,josh2},based on the Gel'fand-Dikii 
identity\cite{gd}, to investigate spectra of such low dimensional field theories. We feel that this method has certain advantages over the inverse scattering method for analysing this type of problems, as it bypasses some of its technical steps. This talk is based on a paper \cite{feinzee} where we have applied this new method to study (\ref{lagrangian}). In particular, our proof that the
Dirac operator in the background of the static solutions of (\ref{saddle}) is
reflectionless is extremely simple, and does not require invoking inverse scattering techniques. It is actually valid for finite $N$. It is worth mentioning at this point that the NJL model (\ref{lagrangian}) is completely integrable for any number of flavors\cite{andrei} and thus its exact spectrum and $S$ matrix are known in detail. The large $N$ spectrum we describe\cite{feinzee,shei} is consistent 
with the relevant exact resuts in \cite{andrei}. However, the powerful methods of \cite{andrei} are inherently limited to $1+1$ dimensional integrable models, 
whereas the analysis\cite{feinzee} we describe here is potentially applicable 
to $1+1$ dimensional non-integrable models as well as to analysing 
inhomogeneous symmetric field configurations in higher dimensions\cite{dhn1}.

\vfill

\pagebreak

\textheight=7.8truein
\renewcommand{\thefootnote}{\alph{footnote}}

\section{Absence of Reflections
in the Dirac Operator With Static Background Fields }
\setcounter{equation}{0}
\renewcommand{\theequation}{2.\arabic{equation}}

\noindent
As was explained in the introduction, we need a manageable form of the diagonal resolvent of the Dirac operator $D\,=\,i\notpa-(\sigx+i\pix\gam_5)$
in a given background of static field configurations $\sigx$ and $\pix$. The extremum condition on $S_{\it eff}$ relates this resolvent, which in principle is a complicated and generally unknown functional
of $\sigx$, $\pix$ and of their derivatives, to $\sigx$ and $\pix$ themselves. This complicated relation is the source of all difficulties that arise in any attempt to solve the model under consideration. It turns out, however, that basic field theoretic considerations, that are unrelated to the extremum condition, imply that $D$ must be reflectionless. This spectral property of $D$ sets rather powerful restrictions on the static background fields $\sigx$ and $\pix$ which are allowed dynamically. In the next section we show how this special property of $D$ allows us to write explicit expressions for the resolvent in some restrictive cases, that are interesting enough from a physical point of view. Inverting $D$ has nothing to do with the large $N$ approximation, and consequently our results in this section are valid for any value of $N$.

The overall energy contained in any relevant static $\si, \pi$ configuration must be finite. Therefore these fields must approach constant vacuum asymptotic values, namely, points on the circle  $\si^2 + \pi^2 \asymptext m^2$, 
with vanishing derivatives. We use the axial $U(1)$ symmetry to fix 
the coordinates in the $\si,\pi$ plane such that $\si (-\infty)=m$ and $\pi (-\infty)=0$. Then $\si (\infty)=m{\rm cos}\theta$ and $\pi (\infty)=m{\rm sin}\theta$, where $\theta$ is the chiral alignment angle of the vacuum at $x=+\infty$ relative to the vacuum at $x=-\infty$.
We use the Majorana representation  $\gam^0=\si_2\;,\; \gam^1=i\si_3$ and 
$\gam^5=-\gam^0\gam^1=\si_1$ for $\gam$ matrices. In this representation  $D$ becomes
\beq
D =\left(\begin{array}{cc} -\pa_x - \si & -i\omega - i\pi \\{}&{}\\ 
i\omega- i\pi & \pa_x - \si\end{array}\right)\,.
\label{dirac1}
\eeq
We invert (\ref{dirac1}) by solving  
\beq
\left(\begin{array}{cc} -\pa_x - \sigx & -i\omega - i\pix \\{}&{}\\ 
i\omega- i\pix & \pa_x - \sigx\end{array}\right)\cdot 
\left(\begin{array}{cc} a(x,y) &  b(x,y) \\{}&{}\\ c(x,y) & 
d(x,y)\end{array}\right)\,=\,-i{\bf 1}\del(x-y)
\label{greens}
\eeq
for the Green's function of (\ref{dirac1}) in a given background $\sigx,\pix$.
By dimensional analysis, we see that the quantities $a,b,c$ and 
$d$ are dimensionles. We analyzed (\ref{greens}) in \cite{feinzee}. The quantities $a(x,y), c(x,y)$ and $d(x,y)$ in (\ref{greens}) may all be inferred from $b(x,y)$, through symmetries of (\ref{dirac1}).  The quantity $b(x,y)$ is  the Green's function of the one dimensional Sturm-Liouville operator
\beq
-\pa_x\left[{\pa_x b(x,y)\over 
\omega+\pix}\right]+\left[\sigx^2+\pix^2-\si'(x)-\omega^2+{\sigx\pi'(x)
\over 
\omega+\pix}\right]{b(x,y)\over \omega+\pix}\,=\,~~\del(x-y)\,.
\label{b}
\eeq

Analysis of the asymptotic behavior of (\ref{greens}) and (\ref{b}) turned out to be quite valuable. The physical asymptotic boundary conditions imposed on $\si,\pi$ are such that the potential term in (\ref{b}) asymptotically a constant, $m^2-\om^2$. We therefore deduced in \cite{feinzee} that the asymptotic behavior of the diagonal resolvent of (\ref{dirac1}) is 
\beqra
\langle x\,|-iD^{-1} | x\,\rangle \asymptotic &&{1 + R\left(k\right)e^{2ik\,|x|}\over 2k}
\left[i\gam_5\pix - \sigx - \gam^0\om\right] 
\nonumber\\{}\nonumber\\
&&+\,{R\left(k\right)e^{2ik\,|x|}\over 2}\,\gam^1~ {\rm sgn}\,x
\label{asymptoticdirac}
\eeqra
where $k=\sqrt{\om^2-m^2}$~ and $R(k)$ is the reflection coefficient
of the Sturm-Liouville operator in (\ref{b}). 
Consider now the expectation value of fermionic vector current\footnote{In the following it is enough 
to discuss only the vector current, because the axial current $j_5^{\mu}=\epsilon^{\mu\nu}j_{\nu}$.} operator $j^{\mu}$ in the static $\sigx,~\pix$ background
\beq 
\langle \sigx,~\pix| j^{\mu} | \sigx, ~\pix\rangle = \int{d\om\over 2\pi} {\rm tr}\left[~\gam^{\mu}~ \langle x\,|-iD^{-1} | x\,\rangle 
\right]\,.
\label{vectorcurrent}
\eeq
We thus find from (\ref{asymptoticdirac}) that the asymptotic behavior 
of the current matrix elements is such that $\langle \sigx,~\pix| j^0 | \sigx, ~\pix\rangle\asymptotic 0$, ~but 
\beq
\langle \sigx,~\pix| j^1 | \sigx, ~\pix\rangle\asymptotic  \int{d\om\over 2\pi}
R\left(k\right)e^{2ik\,|x|}~ {\rm sgn}\,x
\label{currents}
\eeq
where we used the fact that $\int {d\om\over 2\pi} {\om\over k}
f(k) = 0$   because $k\left(\om\right)$ is an even function of $\om$.
Thus, an arbitrary static background $\sigx\,, \pix$ induces space dependent fermion currents that do not decay fast enough as $x\rightarrow\pm\infty$, unless $R(k)\equiv 0$. Clearly, we cannot have such currents in our static problem and we conclude that as far as the field theory (\ref{auxiliary}) is concerned, the fields $\sigx\,, \pix$ must be such that the Sturm-Liouville operators in (\ref{b}) and therefore the Dirac operator (\ref{dirac1}) are reflectionless.

One may also write equations for $\langle \sigx,~\pix| \bar\psi\psi | \sigx, ~\pix\rangle$ 
and\\
$\langle \sigx,~\pix| i\bar\psi\gam_5\psi | \sigx, ~\pix\rangle$ 
that are analogous to (\ref{vectorcurrent}). For reflectionless $\si, \pi$ backgrounds, these equations boil down {\em at the saddle point} (where the expectation value of 
the scalar density is $-\sigx/g^2$, and that of the pseudoscalar density is equal to $-\pix/g^2$) into
\beq\label{spectral}
{1\over Ng^2} = \int{d\om\over 2\pi}~ {1\over \sqrt{\om^2-m^2}}\,.
\eeq
which is simply the gap equation (\ref{bgap}), and thus holds to begin with.

The absence of reflections in (\ref{dirac1}) emerges here from basic principles of field theory, namely, that an initially static $\sigx,~\pix$ will not become time dependnt due to backreaction. Absence of reflections is thus valid beyond the large $N$ saddle point condition from which it was deduced in \cite{dhn,shei}. For reflectionless backgrounds (\ref{asymptoticdirac}) simplifies to
\beq
\langle x\,|-iD^{-1} | x\,\rangle \asymptotic {1\over 2\sqrt{\om^2-m^2}}
\left[i\gam_5\pix - \sigx - \gam^0\om\right] 
\label{asymptotic}
\eeq
This expression has cuts in the complex $\omega$ plane corresponding to scattering states of fermions of mass $m$. These cuts must obviously persist away from the asymptotic region, and we make use of this fact in the next section. 

%\pagebreak

\section{The Diagonal Resolvent for a Fixed Number of Bound States}
\setcounter{equation}{0}
\renewcommand{\theequation}{3.\arabic{equation}}

\noindent
The requirement that the static Dirac operator (\ref{dirac1}) be 
reflectionless is by itself quite restrictive, but in order to actually derive explicit expressions for the resolvent in terms of $\sigx\,,\pix$ and their derivatives we supplement it by assuming in addition that $\sigx$ and $\pix$ 
are such that the spectrum of (\ref{dirac1}) contains a prescribed number of 
bound states. In the following we concentrate on the diagonal resolvent
$B(x)=b(x,x)$ of (\ref{b}). Being the diagonal resolvent of (\ref{b}), $B(x)$ satisfies\cite{feinzee} the Gel'fand-Dikii identity\cite{gd}
\beqra
&&\pax\left\{{1\over \omega + \pix} \pax\left[{\pax B(x)\over 
\omega+\pix}\right]\right\}\nonumber\\{}\nonumber\\ 
&-& {4\over \omega + \pix} \left\{\pax\left[{B(x)\over 
\omega+\pix}\right]\right\}\left[\sigx^2+\pix^2-\si'(x)-\omega^2+
{\sigx\pi'(x)\over\omega+\pix}\right]\nonumber\\{}\nonumber\\ &-& {2B(x)\over 
\left[\omega+\pix\right]^2 }\, \pax\left[\sigx^2+\pix^2-\si'(x) 
+ {\sigx\pi'(x)\over \omega+\pix}\right]\,\equiv\,0\,.
\label{gdblinear}
\eeqra
If we were able to solve (\ref{gdblinear}) for $B(x)$ in a closed form 
for any static configuration of $\sigx, \pix$, we would then be able to 
express $\langle x\,|iD^{-1} | x\,\rangle $ in terms of the latter
fields and their derivatives, and therefore to integrate (\ref{saddle})
back to find an expression for the effective action (\ref{effective}) 
explicitly in terms of $\sigx$ and $\pix$. Invoking at that point Lorentz 
invariance of (\ref{effective}) we would then actually be able to write 
down the full effective action for space-time dependent $\si$ and $\pi$. 
Note moreover that in principle such a procedure would yield an exact 
expression for the effective action, regardless of what $N$ is. 
Unfortunately, deriving such an expression for $B(x)$ in general is a 
difficult task. It becomes manageable only for the special $\si,\pi$ backgrounds specified above.

We now sketch our treatment\cite{feinzee} of a reflectionless background that supports a single bound state. In such a case, $B(x,\om)$ has to have a simple pole on the real $\om$ axis, say at $\om=\om_1$, where clearly, $|\om_1|<m$.  
There is also the continuum cut as in (\ref{asymptotic}). The product of these two factors has dimension $-2$ in mass units. Any other singularity $B(x,\om)$ may have in the complex $\omega$ plane must involve $x$ dependence as well, through the combination ${\rm exp}(i\sqrt{\om^2-m^2}~ x)$. However,
the requirement that the Dirac operator be reflectionless, rules this possibility out. The pole and cut then exhausts all allowed singularities of $B(x,\om)$ in the complex $\om$ plane. In addition, $B(x,\om)$ is dimensionless, and therefore has to be of the form  
\beq
B(x,\om) = { {B_2(x)\omega^{2} + B_{1}(x)\omega +  B_{0}(x)}\over
 {\sqrt{m^2-\omega^2} (\omega-\omega_1)}}\, 
\label{bofx}
\eeq
where the dimension of the unknown function $B_k (x), ~(k=0,1,2) $  is $2-k$. These functions are combinations of $\sigx, \pix$ and their derivatives. Substituting (\ref{bofx}) into (\ref{gdblinear}) results \footnote{Note that because of the linearity and homogeneity of (\ref{gdblinear}),
the purely $\omega$ dependent denominator of (\ref{bofx}) with its
explicit dependence on the bound state energies factors out.} ~in 
a polynomial of degree six in $\omega$, with $x$ dependnet coefficients that 
has to vanish identically. In this way they form an over-determined set of seven differential equations in the five functions $B_2, ..., B_0, \sigx$ and $\pix$.
However, this over-determined system has a solution\cite{feinzee} satisfying the boundary conditions (\ref{asymptotic}),
\beq
B(x) = { \omega  +  \pix \over
2 \sqrt{m^2-\omega^2}} +   {\si^2(x) + \pi^2(x) - \si^{'} (x) - m^2 
\over
4\left(\omega-\omega_1\right) \sqrt{m^2-\omega^2}}\,\quad{\rm where}
\label{bofx1final}
\eeq
\beq
\sigx = m - {m\,{\rm sin} \theta \,{\rm tan} {\theta\over 2} \over 1 + 
{\rm exp}\left[2\om_1 {\rm tan} {\theta\over 2}\cdot 
\left(x-x_0\right)\right]}\quad\quad {\rm and}
\label{sigmapi1bs}
\eeq
\beq
\pix=-[\sigx-m]\,{\rm cot}{\theta\over 2}\,.
\label{pisigma}
\eeq
This linear relation between $\si$ and $\pi$ is not surprising,  because $\si,\pi$ are the two components of an axial vector. Note that the boundary conditions at $x\rightarrow +\infty$ require
\beq
\om_1 \,{\rm tan} {\theta\over 2} < 0\,.
\label{negative}
\eeq
Our results (\ref{sigmapi1bs}) for $\sigx$ and $\pix$ agree with those of \cite{shei}. They have the profile of an extended object, a lump or a chiral ``bag",  of size of the order ${\rm cot} {\theta\over 2}/\om_1$ 
centered around an arbitrary point $x_0$. Note that the profiles in (\ref{sigmapi1bs}) satisfy 
\beq
\rho^2(x) = \si^2(x) + \pi^2(x) = m^2 - m^2\,{\rm sin}^2\,(\theta/2)\,{\rm sech}^2\,\left[\om_1 {\rm tan} 
{\theta\over 2} \cdot\left(x-x_0\right)\right]\,.
\label{rho1bs}
\eeq
Thus, as expected by construction, this configuration interpolates between 
two different vacua at $x=\pm\infty$. As $x$ increases from $-\infty$, the vacuum configuration becomes distorted. The distortion reaches its maximum at the location of the ``bag", where $m^2-\rho^2(x_0) =  m^2\,{\rm sin}^2\,(\theta/2)$ and then relaxes back into the other vacuum state at $x=\infty$.  The arbitrariness of $x_0$ is, of course, a manifestation of translational invariance.

In \cite{feinzee} we also have partial results for the case of two bound 
states, at energies $\om_1, \om_2$. In particular we found that $y(x)=\sigx-m$ satisfies the differential equation
\beqra
&&2\,\la\,(\om_1+\om_2)\,[4m\,y^2 + 2(1+\la^2) y^3 - y'']
\nonumber\\
&&+\pax\{\,4\,(m^2+\om_1\om_2)\,y + 6\,m\,y^2 + 2\,(\la^2+1)\,y^3 - y''\} = 0  
\label{sieq2}
\eeqra
where $\la = -{\rm cot} (\theta/2)$, and $\pix$ is given by (\ref{pisigma}).
Note that if we set $\om_1+\om_2=0$ and $\pix = 0$ the spectrum becomes invariant under $\om\rightarrow -\om$, and we obtain the equation appropriate to the Gross-Neveu model.

\pagebreak

\section{The Saddle Point Conditions}
\setcounter{equation}{0}
\renewcommand{\theequation}{4.\arabic{equation}}

\noindent

Derivation of the explicit expressions of $\sigx$ and $\pix$ does not involve the saddle point equations (\ref{saddle}). 
Rather, it tells us independently of the large $N$ approximation that $\sigx$ and $\pix$ must have the form given in 
(\ref{sigmapi1bs}) in order for the associated Dirac 
operator to be reflectionless and to have a single bound state at a 
prescribed energy $\om_1$ in addition to scattering states. Thus, for the solution (\ref{sigmapi1bs}) we have yet to determine the values of $\om_1$
and $\theta$ allowed by the saddle point condition (\ref{saddle}). It is this dynamical feature that we can analyse only in the large $N$ limit. Note that 
the from (\ref{bofx1final}) we can reconstruct the other three entries 
of $\langle x\,|-iD^{-1} | x\,\rangle$. Substituting this resolvent, with $\si, \pi$ given by (\ref{sigmapi1bs}) into (\ref{saddle}) we obtain a quantization 
condition for $\om_1$, in the form of coupled dispersion integrals that have to vanish.  A nice feature of these integrals is that their potentialy 
ultra-violet divergent parts are equal to the difference of the two sides of 
the gap equation (\ref{spectral}), and are therefore free of such divergences.
Some $\om$ poles in these integrals are at $x$ dependent locations. 
The quantization condition on $\om_1$ cannot be $x$ 
dependent, so the residues of these poles must vanish. This happens provided 
the parameters in (\ref{sigmapi1bs}) satisfy
\beq
\om_1^2 = m^2 {\rm cos}^2 ({\theta\over 2})
\label{luckily}
\eeq
This relation actually leaves $\theta$ the only free parameter in the problem with respect to which we have to extremize the action. The condition (\ref{negative}) then picks out one branch of (\ref{luckily}). Assuming that the Dirac sea is completely filled, and that the bound state traps $n<N$ fermions,
the saddle point condition becomes equivalent\cite{feinzee} to the requirement that $S_{eff}$ evaluated at the appropriate $\si,\pi$ configurations, be
stationary as a function of $\theta$, namely, 
\beq
{1\over NT}{\pa S\over \pa\theta} = {m\over 2}\left({n\over N} + {\theta\over 
2\pi}\right) {\rm sin}{\theta\over 2}\,.
\label{nigzeret}
\eeq
Only the critical value at 
\beq
\theta = - {2\pi n \over N}\,,
\label{theta}
\eeq
corresponds to extended objects. The other zeros of (\ref{nigzeret}) at $\theta_k = 2\pi k,~ k\in {\bf Z}$  do not correspond to such objects at all 
as should be clear from (\ref{sigmapi1bs}). We therefore discard them, and concentrate on the extremum given by (\ref{theta}). The bound state energy is therefore $\om_1=m{\rm cos}\left({n\pi\over N}\right)$. Equation (\ref{theta}) asserts that the relative chiral rotation of the vacua at $\pm\infty$ is proportional to the number of fermions trapped in the ``bag". Our result (\ref{theta}) is consistent with 
the fact that the fermion number current in a soliton background can be  determined in some cases by topological considerations\cite{goldstone}. Note from (\ref{theta}), that in the large $N$ limit, $\theta$ (and therefore $\om_1$)
take on non-trivial values only when the number of the trapped fermions scales
as a finite fracion of $N$.

As we already mentioned in the introduction, ``bags" 
formed in the NJL model are not stable because of topology. They are 
stabilized by releasing binding energy of the fermions trapped in them.
To see this more explicitly, we calculate now the mass of the ``bag" corresponding to (\ref{sigmapi1bs}) and (\ref{theta}). Integrating (\ref{nigzeret}) with respect to $\theta$ we find
\beq
{-1\over NTm} S(\theta)  = \left({n\over N} + {\theta\over 
2\pi}\right) {\rm cos}{\theta\over 2} - {1\over \pi}{\rm sin}{\theta\over 2}\,.
\label{stheta}
\eeq
Note that (\ref{stheta}) is not manifestly periodic in $\theta$ because the Pauli exclusion principle limits $\theta$ to be between $0$ and $2\pi$.
The mass of a ``bag" containing $n$ fermions in a single bound state
is given by the energy $E(\theta)=-S(\theta)/T$ evaluated at the appropriate chiral angle (\ref{theta}). We thus find that this mass is simply
\beq
M_n={Nm\over\pi} {\rm sin}{\pi n\over N}
\label{bagmass}
\eeq
in accordance with \cite{shei,andrei}. It is easy to check that (\ref{bagmass}) is a minimum of $E(\theta)$ for $0<{n\over N}<1$. These ``bags" are stable because 
\beq
{\rm sin}{\pi \left(n_1 + n_2\right)\over N} < {\rm sin}{\pi n_1\over N}\,+\,{\rm sin}{\pi n_2\over N}
\label{stablebag}
\eeq
for $n_1+n_2$ less than $N$, such that a ``bag" with $n_1+n_2$ fermions 
cannot decay into two ``bags" each containing a lower number of fermions.

Entrapment of a small number of fermions cannot distort the homogeneous vacuum considerably, so we expect that $M_n$ will be roughly the mass of 
$n$ free massive fermions for $n<<N$ as (\ref{bagmass}) indeed shows. As a matter of fact we used this expectation to fix the integration constant in (\ref{stheta}). However, as the number of fermions trapped in the ``bag" approaches $N$, $M_n$ vanishes and the fermions release
practically all their rest mass $Nm$ as binding energy, to achieve maximum stability\cite{mackenzie}. Note from (\ref{theta}), that the soliton twists all the way around as the number of fermions approaches $N$. 

We conclude this talk making some remarks on the case of two bound states.
In \cite{feinzee} we found that the bound state energies must be located at 
$\om_1 = m\, {\rm cos}\left(n_1\pi\over N\right)\,,\quad \om_2 = m \,{\rm cos}\left(n_2\pi\over N\right)$ which are identical in form to single bound state energy levels. From the general considerations of \cite{goldstone} we expect that the chiral angle $\theta$ will be proportional to the total number of fermions trapped by the ``bag", so (\ref{theta}) must read now
$\theta = - {2\pi \left(n_1 + n_2\right) \over N}\,$. Clearly, the mass of the ``bag" will depend on $n_1$ and $n_2$ seperately, and not only on their sum through $\theta$.

\pagebreak
\nonumsection{Acknowledgements}
\noindent
JF would like to thank the organizers for an interesting and  stimulating workshop. This work was partly supported by the National 
Science Foundation under Grant
No. PHY89-04035. 

\nonumsection{References}


\noindent
\begin{thebibliography}{000}
\bibitem{njl} Y. Nambu and G. Jona-Lasinio, Phys. Rev. {\bf 122}, 345 (1961), {\it ibid} {\bf 124}, 246 (1961).
\bibitem{gn}  D.J. Gross and A. Neveu,  Phys. Rev. D  {\bf 10},   3235
(1974).
\bibitem{feinzee} J. Feinberg and A. Zee, Preprint NSF-ITP-96-15, March 1996
(cond-mat/9603173).
\bibitem{chirallagrangian} For a recent review on the NJL model as an effective low energy description of QCD in four space time dimensions see
J. Bijens, Phys. Rep. {\bf 265}, 369 (1996).\\
See also R.T. Cahill and C.D. Roberts, Phys. Rev. D {\bf 32}, 2419 (1985).
\bibitem{shei} S. Shei,  Phys. Rev. D {\bf 14}, 535 (1976).
\bibitem{sphericalbag} T. D. Lee and G. Wick, Phys. Rev. D {\bf 9}, 2291 (1974);
~R. Friedberg, T.D. Lee and R. Sirlin, Phys. Rev. D {\bf 13}, 2739 (1976);
~R. Friedberg and T.D. Lee, Phys. Rev. D {\bf 15}, 1694 (1976), {\em ibid.} 
{\bf 16}, 1096 (1977);
~A. Chodos, R. Jaffe, K. Johnson, C. Thorn, and V. Weisskopf,  Phys. Rev. D {\bf 9}, 3471 (1974).
\bibitem{shellbag} W. A. Bardeen, M. S. Chanowitz, S. D. Drell, M. Weinstein 
and T. M. Yan, Phys. Rev. D {\bf 11}, 1094 (1974);~M. Creutz, Phys. Rev. D {\bf 10}, 1749 (1974).
\bibitem{decouple}
See the concluding remarks in \cite{shei} who brings an unpublished argument
for decoupling of $\theta$ due to R. Dashen, along the lines of M.B. Halpern, Phys. Rev. D {\bf 12} 1684 (1975).~See also E. Witten,  Nucl. Phys. {\bf B145}, 110 (1978).
\bibitem{coleman} S. Coleman, Commun. Math. Phys. {\bf 31}, 259 (1973). 
\bibitem{mackenzie} R. MacKenzie, F. Wilczek and A.Zee,  Phys. Rev. Lett {\bf 53}, 2203 (1984).
\bibitem{ccgz} C.G. Callan, S. Coleman, D.J. Gross and A. Zee,
unpublished;~D.J. Gross in {\sl Methods in Field Theory\/}, R. Balian and J. Zinn-Justin (Eds.), Les-Houches session  XXVIII 1975 (North Holland, Amsterdam, 1976);~A. Klein, Phys. Rev. D {\bf 14}, 558 (1976);~See also \cite{josh1} and R. Pausch, M. Thies and V. L. Dolman, Z. Phys. A {\bf 338}, 441 (1991).
\bibitem{dhn}  R.F. Dashen, B. Hasslacher and A. Neveu,  Phys. Rev. D 
{\bf 12},2443 (1975).
\bibitem{josh1} J. Feinberg,  Phys. Rev. D {\bf 51}, 4503 (1995).
\bibitem{faddeev}  L.D. Faddeev and L.A. Takhtajan, {\sl Hamiltonian Methods
in the Theory of Solitons\/} (Springer Verlag, Berlin, 1987). 
\bibitem{josh2} J. Feinberg, Nucl. Phys. {\bf B433}, 625 (1995).
\bibitem{gd} I.M. Gel'fand  and L.A. Dikii, Russian Math. Surveys~~{\bf
30},~77{}~(1975).
\bibitem{andrei} N. Andrei and J.H. Lowenstein,  Phys. Rev. Lett. {\bf 43}, 
1698 (1979),~ Phys. Lett. {\bf 90B}, 106 (1980), {\it ibid.} {\bf 91B} 401 (1980).  
\bibitem{dhn1}  R.F. Dashen, B. Hasslacher and A. Neveu,  Phys. Rev. D 
{\bf 10}, 4130 (1974).
\bibitem{goldstone} J. Goldstone and F. Wilczek, Phys. Rev. Lett. {\bf 47}, 
986 (1981);~See also Eq. ($4.44$) of R. Aviv and A. Zee, Phys. Rev. D. {\bf 5}, 
2372 (1972).
\end{thebibliography}
\end{document}